\documentclass[runningheads,a4paper]{llncs}

\usepackage{amssymb}
\usepackage{graphicx}

\usepackage{url}
\newcommand{\keywords}[1]{\par\addvspace\baselineskip
\noindent\keywordname\enspace\ignorespaces#1}

\begin{document}

\mainmatter  % start of an individual contribution

\title{Distributed Generation and Resilience in Power Grids}

\titlerunning{Distributed Generation and Resilience in Power Grids}

\author{Antonio Scala\inst{1,2,3}
%\email{antonio.scala@phys.uniroma1.it}
%\thanks{Please note that the LNCS Editorial assumes that all authors have used the western naming convention, with given names preceding surnames. This determines the structure of the names in the running heads and the author index.}%
\and Mario Mureddu\inst{4,5} \and \\
Alessandro Chessa\inst{1,5} \and Guido Caldarelli\inst{2,1,3} \and
Alfonso Damiano\inst{6} }
\authorrunning{Antonio Scala et al.}

\institute{
ISC-CNR Physics Dept., Univ. "La Sapienza" Piazzale Moro 5, 00185 Roma, Italy
\and IMT Alti Studi Lucca, piazza S. Ponziano 6, 55100 Lucca, Italy
\and London Institute of Mathematical Sciences, 22 South Audley St
\\Mayfair London W1K 2NY, UK
\and Department of Physics, University of Cagliari, Italy 
\and Linkalab, Complex Systems Computational Laboratory, 09129 Cagliari, Italy
\and Dipartimento di Ingegneria Elettrica ed Elettronica
 Universit\`a di Cagliari, Italy
%\url{http://www.springer.com/lncs}
}


%\toctitle{Lecture Notes in Computer Science}
%\tocauthor{Authors' Instructions}
\maketitle

\begin{abstract}
We study the effects of the allocation of distributed generation
on the resilience of power grids. We find that an unconstrained allocation
and growth of the distributed generation can drive a power grid beyond
its design parameters. In order to overcome such a problem, we propose
a topological algorithm derived from the field of Complex Networks
to allocate distributed generation sources in an existing power grid.
\keywords{distributed generation, AC power model, complex networks, pagerank}
\end{abstract}

\section{Introduction}

Distributed Generation from renewable sources is having a deep
impact on our power grids. The difficult task of integrating the
stochastic and often volatile renewable sources into a the grid
designed with a power-on-demand paradigm could perhaps solved
leveraging on distributed storage \cite{Baghaie2010}; nevertheless,
massive and economic power storage is not yet readily available. In
the meanwhile, power grids are nowadays required to be robust and
smart, i.e. systems able to maintain, under normal or perturbed
conditions, the frequency and  amplitude variations of the supplied voltage 
into a defined range and to provide fast restoration after faults.
Therefore, many studies have concentrated on the dynamic behaviour
of Smart Grids to understand how to ensure stability and avoid loss
of synchronization during typical events like the interconnection of
distributed generation. The large number of elements present into
real grids calls for simplifications like the mapping among the
classic swing equations \cite{StaggBOOK1968} and Kuramoto models
\cite{Filatrella2008,Fioriti2009,DorflerSIAM2012} that allows
to study numerically or analytically the synchronization and the
transient stability of large power networks.

Even simple models \cite{DobsonHICSS2001} akin to the DC power flow
model \cite{WoodBook1984} show that the network topology can dynamically
induce a complex size probability distributions of blackouts (power-law
distributed), both when the system is operated near its
limits \cite{CarrerasCHAOS2002}
or when the system is subject to erratic disturbances \cite{SachtjenPRE2000}.
New realistic metrics to assess the robustness of the electric power grid
with respect to the cascading failures \cite{YoussefITCP2011} are therefore needed.

Smart grids are going to insist on pre-existing networks designed
for different purposes and tailored on different paradigms and new
kind of failures are possible: therefore a careful transition is needed.
One possible approach could be the use of advanced metering infrastructure
(AMI) not only for implementing providers and customers services,
but also to detect and forecast failures; nevertheless an ill-designed
network will never be efficient.

Our approach will not concentrate on the instabilities but will focus instead
on the condition under which, in presence of distributed generation,
the system can either be operated or controlled back within its design parameters,
i.e. it is \emph{resilient}. It is akin in spirit to the approach
of \cite{Chertkov2011}, that by applying DC power flow analysis
to a system with a stochastic distribution of demands,
aims to understand and prevent failures by identifying
the most relevant load configurations on the feasibility boundary
between the normal and problematic regions of grid operation.

To model power grids, we will use the more computational
intensive AC power flow algorithms as, although DC flows are on average
wrong by a few percent \cite{Stott2009}, error outliers could distort
our analysis.

To model distributed renewable sources, we will introduced a skewed
probability distribution of load demands representing a crude
model of reality that ignores the effects like the correlations 
(due for examples to weather conditions) between
different consumers or distributed producers .

\section{Methods}

\subsection{AC Power Flow}

The AC power flow is described by a system of non-linear equations
that allow to obtain complete voltage angle and magnitude information
for each bus in a power system for specified loads \cite{GraingerBOOK1994}.
A bus of the system is either classified as Load Bus if there are
no generators connected or as a Generator Bus if one or more generators
are connected. It is assumed that the real power $PD$ and the reactive
power $QD$ at each Load Bus are given, while for Generator Buses the
real generated power $PG$ and the voltage magnitude $|V|$ are given.
A particular Generator Bus, called the Slack Bus, is assumed as a
reference and its voltage magnitude $|V|$ and voltage phase $\Theta$
are fixed. The branches of the electrical system are described by
the bus admittance matrix $Y$ with complex elements $Y_{ij}$s.

The power balance equations can be written for real and reactive power
for each bus. The real power balance equation is:

\[
0=-P_{i}+{\displaystyle \sum_{k=1}^{N}\left|V_{i}\right|}\left|V_{k}\right|\left(G_{ik}\cos\theta_{ik}+B_{ik}\sin\theta_{ik}\right)
\]

where $N$ is the number of buses,
$P_{i}$ is the net real power injected at the $i^{th}$ bus
, $G_{ik}$ is the real part and $B_{ik}$ is the imaginary part of
the element $Y_{ij}$ and $\theta_{ik}$ is the difference in voltage
angle between the $i^{th}$ and $k^{th}$ buses. The reactive power
balance equation is:

\[
0=-Q_{i}+{\displaystyle \sum_{k=1}^{N}\left|V_{i}\right|}\left|V_{k}\right|\left(G_{ik}\sin\theta_{ik}-B_{ik}\cos\theta_{ik}\right)
\]

where $Q_{i}$ is the net reactive power injected at the $i^{th}$
bus.

Real and reactive power flow on each branch as well as generator reactive
power; the output can be analytically determined but due to the non-linear
character of the system numerical methods are employed to obtain a
solution. To solve such equations, we employ Pylon \cite{pylon},
a port of MATPOWER \cite{matpower} to the Python programming language.

A requirement for the stability of the load and generation requirements
is the condition that all branches and buses operate within their
physical feasibility parameters; going beyond such parameters can
trigger cascades of failures eventually leading to black outs \cite{Pahwa2010}.

In the present paper a topological investigation on the power grid
has been developed in order to evaluate the effects of distributed
generation on the voltage and power quality. Hence, a steady state
analysis has been carried out and the transient phenomena connected
to the power flow control have been neglected. Under this hypothesis
the frequency variation connected to power flow control has been
considered stabilized and the system has been considered
characterized  by a constant steady state supply voltage frequency.
Therefore, if all the nodes are near their nominal voltage, it is
much easier to control the system and to avoid reaching infeasible
levels of power flow. Consequently, to measure the effects of power
quality of a power grid under distributed generation we measure the
fraction $F$ of load buses whose tension goes beyond $\pm5\%$ of its
nominal voltage. Notice that real networks are often operated with
some of the buses beyond such parameters so that (especially for
large networks) it is expected to be $F\neq0$ under operating conditions. 
The maximum of the resilience for a power grid (intended as the capability 
of restoring full feasible flows) is expected to be for $F=0$.

\subsection{Distributed Generation and Skew-normal distribution}

We will consider distributed generation due to erratic renewable sources
like sun and wind; therefore, we will model the effects of ``green
generators'' on a power grid as a stochastic variation the power
requested by load buses. Load buses with a green generator will henceforth
called green buses. We will consider the location of green buses to
be random; the fraction $p$ of green buses will characterize the
penetration of the distributed generation in a grid.

If the power dispatched by distributed generation is high enough,
loads can eventually become negative: this effect can be related to
the efficiency of green generators. We model such an effect by considering
the load on green buses described by the skew-normal distribution
\cite{Azzalini2005}, a pseudo-normal distribution with a non-zero
skewness:
\[
f\left(x,\alpha\right)=2\phi\left(x\right)\Phi\left(\alpha x\right)
\]
where $\alpha$ is a real parameter and
\[
\begin{array}{ccc}
\phi\left(x\right)=\exp\left(-x^{2}/2\right)/\sqrt{2\pi}\,\,\,\,\,\, &  & \,\,\,\,\,\,\Phi\left(\alpha x\right)=\int_{-\infty}^{\alpha x}\phi\left(t\right)dt\end{array}
\]
The parameter $\alpha$ will characterize the level of the
distributed generation: to positive $\alpha$ correspond loads
positive on average, while for negative $\alpha$ green nodes will
tend to dispatch power.

Our model grids will therefore consist of three kind of buses: $N_G$ generators
(fixed voltage), $N_l$ pure loads (fixed power consumption) and $N_g$ green buses
(stochastic power consumption) with $N_G+N_l+N_g=N$ the total number of buses and
$N_g+N_l=N_L$ the number of load nodes.
The fraction $p=N_g/N_L$ measures the penetration of renewable
sources in the grid.

\subsection{Complex Networks and Page Rank}

The topology of a power grid can be represented as a directed graph
$G=\left(V,E\right)$, where to the $i$-th bus corresponds the nodes
$n_{i}$ of the set $V$ and to the $k$-th branch from the $i$-th to the
$j$-th bus corresponds the edge $e_k=\left(i,j\right)$ of the set $E$. In Power
System engineering, it is custom to associate to the graph $G$ representing
a power networks its \emph{incidence} matrix $B$ whose elements are
\[
B_{ik}=\left\{ \begin{array}{ccc}
&1 & if\,e_k=\left(i,\_\right) \in E\\
-&1 & if\,e_k=\left(\_,i\right) \in E\\
&0 & otherwise
\end{array}\right.
\].
An alternative representation of the graph much more used in other scientific fields is its \emph{adjacency}
matrix $A$ whose element are
\[
A_{ij}=\left\{ \begin{array}{cc}
1 & if\,\left(i,j\right) \in E\\
0 & otherwise
\end{array}\right.
\]
%$A_{ij}$ is $1$ if there is a link from
%$n_{i}$ to $n_{j}$, $0$ otherwise.

While Graph Theory has an old tradition since Euler's venerable problem
on Koenigsberg bridges \cite{BergeBOOK1958}, Complex Networks is the new field
investigating the emergent properties of large graphs. An important
characteristic of the nodes of a complex network is their centrality,
i.e. their relative importance respect to the other nodes of the graph
\cite{CaldarelliBOOK2007}. An important centrality measure is Page
Rank, the algorithm introduced Brin and Page \cite{Brin1998} to rank web pages
that is at the hearth of the Google search engine. The Page Rank $r_{i}$
of the $i$-th node is the solution of the linear system
\[
r_{i}=\frac{1-\rho}{N}+\rho{\displaystyle \sum\frac{A_{ij}r_{j}}{d_{j}^{o}}}
\]
where $N$ is the number of buses (nodes), $d_{i}^{o}={\displaystyle \sum_{i}A_{ij}}$
is the number of outgoing links (out-degree) and $\rho=0.85$ is the
Page Rank damping factor. In studying power grids, we will employ
Page Rank as it is strictly related to several invariants occurring
in the study of random walks and electrical networks \cite{ChungPRandRW}.

\section{Results}

\subsection{Effects of distributed generation}

\begin{figure}
\centering
\includegraphics[height=6.2cm]{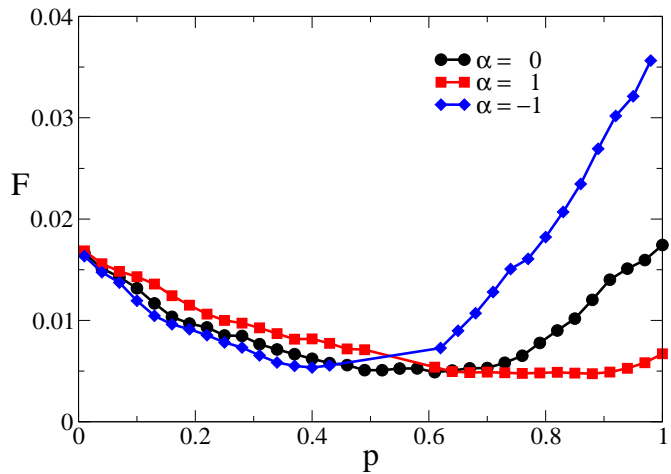}
\caption{Effects of the penetration $p$ of distributed generation on
the resilience of the Polish power grid at different values $\alpha$
of the green generators. Notice that for $\alpha=0$ renewable
sources satisfy on average the load requested by the network, while
for $\alpha<0$ there is a surplus of renewable energy. Lower values
of the fraction $F$ of buses operating near their nominal tension
correspond to a higher resiliency. Notice that the penetration of
distributed generation initially enhances resiliency. At higher
values of $p$, resilience worsens; in particular, it is severely
impaired if distributed generation produces on average more energy
than the normal load requests ($\alpha=-1$). It is therefore
advisable to keep levels of renewable energy production below
the normal load request ($\alpha=1$).} \label{fig:alpharandom}
\end{figure}

We have investigated the effects of our null model of distributed
generation on the 2383 bus power grid of Poland, 1999. Starting from
the unperturbed network, we have found an initial fraction
$F_{0}\cong1.6\%$ of load buses beyond their nominal tension. We
have therefore varied the penetration $p$ at fixed distributed
generation level  $\alpha$'s; results are shown in Fig.
\ref{fig:alpharandom}.
\\
\\

We find that the behaviour of the fraction $F$ of buses operating near
their nominal tension does not follow a monotonic behaviour. Initially
(low values of $p$), the penetration of distributed generation enhances
resiliency (i.e. decreases $F$). At higher values of $p$, $F$ grows
and resilience worsens. Such an effect is particularly severe if green
nodes introduce a surplus ($\alpha<0$) of power respect to the normal
($p=0$) operating load requests. On the other hand, keeping the levels
of renewable energy production below ($\alpha>0$) the normal load
request delays the point beyond which the penetration of distributed
generation worsens the resiliency.

Notice that when distributed generation is ancillary ($\alpha>0$) and not 
predominant in the power supplied of the network, full penetration ($p=1$)
of renewable sources lead to more stable state than
the initial ($p=0$) one.

\subsection{Targeted distributed generation}

\begin{figure}
\centering
\includegraphics[height=12.4cm]{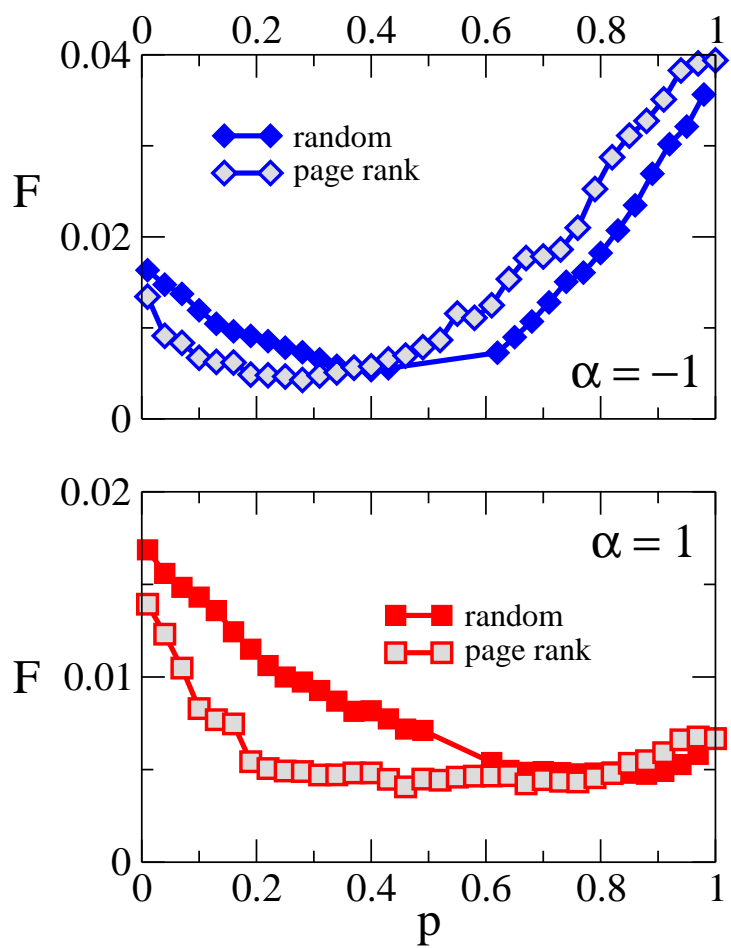}
\caption{Comparison between random placement (filled symbols) and page-rank placement (empty symbols) of green generators in the Polish grid, both for surplus production of renewable energy (upper panel, $\alpha=-1$) and for levels of renewable energy
production below the normal load request (lower panel, $\alpha=1$).
The page-rank placement of renewable sources allows to attain lower values of the fraction $F$ of buses operating near their nominal tension (and hence a higher resiliency) at lower values of the penetration $p$. The best case is realized for levels 
of renewable energy production below the normal load request, where a plateau 
to low values of $F$ is quickly attained.}
\label{fig:alphapagerank}
\end{figure}

Beside their natural application to web crawling, the Page Rank algorithm can 
be applied to find local partitions of a network that optimize conductance
\cite{Andersen2006}. We therefore investigate what happens in a power network if
distributed generation is introduced with a policy  that accounts for the pagerank
of load nodes. In other words, for a level of penetration $p$, we choose the
first $n_g=pN_L$ load nodes in decreasing pagerank order to become green nodes.
The effects of such a choice are shown compared to the random penetration policy
in Fig. \ref{fig:alphapagerank}.

We find that, for low penetration levels, the pagerank policy reduces the
number of nodes operating beyond their nominal tension both for positive and for
negative $\alpha$'s. Again, the excess of power production ($\alpha<0$) comparatively 
reduces the resilience of the network.

Preliminary results show that Page Rank is the best behaved among centralities
in enhancing  power grid resilience; such study will be the subject of a
future publication.

\section{Discussion}

We have introduced a model base on the AC power flow equation that allows to
account for the presence of erratic renewable sources distributed on a power grid
and for their efficiency.
By defining the resilience of the grid as a quantity related to the possibility of
controlling the power flow via voltage adjustments (hence returning within the
operating bounds of its components), we have studied the penetration of
distributed generation on a realistic power grid.

We have found that while the introduction of few "green" generators
in general enhances the resilience of the network by decreasing the
number of nodes operating beyond their nominal voltage, a further
increase of renewable sources could decrease the power quality of the
grid. Anyhow, if distributed generation is ancillary and not
predominant in the power supplied of the network, the grid at full
penetration ($p=1$) of renewable sources is in a more stable
state than the starting grid ($p=0$).

Our finding that a surplus of production from renewable
sources is also a source of additional instabilities is perhaps
to be expected in general for networks that have been designed to dispatch power 
from their generators to their loads and not to produce energy "locally".
While we have found that in an isolated grid 
instability possible increases with the penetration, 
what happens when more grids are linked together is an open subject.
Power grids are typical complex infrastructural systems; therefore they
can exhibit emergent characteristics when they interact with each other,
modifying the risk of failure in the individual systems \cite{CarrerasHICSS2007}.
As an example, the increase in infrastructural interdependencies
could either mitigate \cite{BrummittPNAS2012}
or increase \cite{LaprieCORR2008,BuldyrevNAT2010} the risk of a system failure.

Finally, we find that a policy of choosing the sites where to introduce
renewable sources according to Page Rank allows to increase
the resilience with a minimal amount of green buses. Such policy does not 
take into account other factors and should therefore be integrated in a 
multi-objective optimization to consider the environmental, economical and 
social constraints.

\subsubsection*{Acknowledgements.}
We thank US grant HDTRA1-11-1-0048, EU FET Open project FOC nr.255987 and CNR-PNR National Project ”Crisis-Lab” for support. 
The contents do not necessarily reflect the position or the policy of funding parties.

\bibliography{critis2012}

\end{document}